# Change of caged dynamics at $T_g$ in hydrated proteins found after suppressing the methyl-group rotation contribution

K. L. Ngai[a][1], S. Capaccioli[a,b][2], A. Paciaroni[c][3]

[a]*CNR-IPCF, Dipartimento di Fisica, Largo Bruno Pontecorvo 3,I-56127, Pisa, Italy*
[b]*Dipartimento di Fisica, Università di Pisa, Largo Bruno Pontecorvo 3 ,I-56127, Pisa, Italy*
[c]*Dipartimento di Fisica, Università di Perugia & IOM-CNR, , Via A. Pascoli 1, 06123 Perugia, Italy*

___

**Abstract**

In conventional glassformers at sufficiently short times and low enough temperatures, molecules are mutually caged by the intermolecular potential. The fluctuation and dissipation from motion of caged molecules when observed by elastic incoherent neutron scattering exhibit a change in temperature dependence of either the elastic intensity or the mean square displacement (MSD) at the glass transition temperature $T_g$. This is a general and fundamental property of caged dynamics in glassformers, which is observed always at or near $T_g$ independent of the energy resolution of the spectrometer. Recently we showed the same change of temperature dependence at $T_g$ is present in proteins solvated with bioprotectants, such as sugars or glycerol with or without the addition of water, coexisting with the dynamic transition at a higher temperature $T_d$. These solvated proteins all have higher $T_g$ and $T_d$ than the proteins hydrated by water alone, and thus the observation of the change of $T$-dependence of the MSD at $T_g$ is unobstructed by the methyl-group rotation contribution at lower temperatures and the dynamic transition at higher temperatures. On the other hand, proteins hydrated by water alone have lower $T_g$ and $T_d$, and hence unambiguous evidence of the change of the $T$-dependence of the MSD at $T_g$ is hard to find. Notwithstanding, evidence on the break of the MSD at $T_g$ can be found on manipulating the sample by either fully or partially deuterating the protein to suppress the methyl-group rotation contribution. An alternative strategy is the use of a spectrometer that senses motions with characteristic times shorter than ~15 ps, which confers the benefit of shifting both the onset of methyl-group rotation contribution as well as the dynamic transition to higher temperatures, and again the change in $T$-dependence of the MSD at $T_g$ becomes evident. The break of the elastic intensity or the MSD at $T_g$ coexists with the dynamics transition at $T_d$ in hydrated and solvated proteins. Recognition of this fact helps to remove inconsistency and conundrum encountered in interpreting the data that thwart progress in understanding the origin of the dynamic transition and its connection to biological function.

---

[1] ngai@df.unipi.it, Tel. +39-0502214322, Fax. +39-0502214333
[2] capacci@df.unipi.it Tel. +39-0502214537, Fax. +39-0502214333
[3] alessandro.paciaroni@fisica.unipg.it, Tel. +39-0755852785

**Introduction**

The dynamics of hydrated proteins at low temperatures are principally determined by harmonic vibrations. At higher temperatures deviation from harmonic behavior is found due to the onset of relaxation processes or non-vibrational motion. The deviation is hydration level dependent, which gives rise to mobility or flexibility of the protein to perform its biological function. This change in dynamics was first found by Mössbauer spectroscopy in hydrated myoglobin as a distinct crossover from linear $T$-dependence at lower temperatures to a stronger temperature dependence of the mean-square-displacement (MSD) of the heme iron at some temperature.[1,2,3]. Mössbauer spectroscopy probes motion on the timescale of 140 ns. This change of dynamics of myoglobin hydrated with $D_2O$ was found also by inelastic neutron scattering in a shorter timescale from 0.1 to 100 ps[4]. The same onset of anharmonic motions was found in other hydrated proteins [5-22], and the effect is referred to in the literature as the protein dynamical transition at temperature $T_d$. It is by now generally accepted that $T_d$ depends on the energy resolution and the corresponding time window of the neutron scattering spectrometer used [23].

In the original neutron scattering experiment on the dynamical transition in myoglobin hydrated with $D_2O$, protein fluctuations were measured as a function of temperature $T$ using the IN13 backscattering spectrometer with energy resolution of 8 µeV and timescale within the range from 4 to 100 ps [4]. The MSD, $<u^2(T)>$, starts to deviate from linear $T$-dependence of the contribution from harmonic vibration near and below 150 K. This first break of $<u^2(T)>$ from harmonic behavior is now recognized as coming from methyl group rotation. After the first break, $<u^2(T)>$ continues to increase mildly with temperature until about 240 K, after which begins a more rapid increase with temperature. This second break is identified with the dynamic transition at $T_d$~240 K [24]. Subsequent experiments on other hydrated proteins have confirmed the observation of the first break from harmonic behavior at lower temperature is due to the rotational relaxation times of the methyl groups in the protein entering the time window of the spectrometer used, although the onset temperature $T_{on}$ depends on the protein and can range from 100 to 150 K [5,15,16,25-27]. The methyl group rotation contribution, and hence the first break, is absent if the hydrated biomolecule[28] or homomeric polypeptide [20] has no methyl group, or if the protein is deuterated and the dynamic transition is probed via the hydrated water [14,26,29].

Most hydrated proteins are also glass-formers. Glass transition at temperature $T_g$ can be detected either by calorimetry [30-33] or by dielectric [34] or mechanical [35,36] relaxation spectroscopy. The glass transition and the dynamic transition are two different processes in the same solvated or hydrated protein, and $T_d$ is higher than $T_g$, in general.

In a paper preceding the present one [37], we have shown in several proteins, solvated with sugars or glycerol with or without the addition of water, the presence of another break at $T_g$ in the temperature dependence of $<u^2(T)>$ from neutron scattering in addition to the familiar break due to the dynamic transition at $T_d$. This was done by reexamination of published as well as previously unpublished neutron scattering data of some hydrated and solvated proteins. The new break occurs near $T_g$ which lies in between $T_{on}$ and $T_d$, and it exhibits a change from weaker temperature dependence of $<u^2(T)>$ to a stronger one on crossing $T_g$ from below. The same break is found in the temperature dependence of $<u^2(T)>$ at $T_g$ of the pure solvent without protein also by neutron scattering using spectrometer with the same resolution. This finding in the pure solvent is unsurprising if we recall the change of $T$-dependence of $<u^2(T)>$ on crossing $T_g$ is actually a general property of all glassformers [38], and the solvents are glassformers. It is found not only from $<u^2(T)>$ by neutron scattering but also from $T$-dependence of susceptibility loss at fixed frequency by dynamic light scattering in the GHz frequency range [39] and dielectric relaxation at lower frequencies and correspondingly lower temperatures [40]. As a general property of glassformers, it has been considered as one of the important problems to be solved in the research field of glass transition [38,41]. The discovery of the new break at $T_g$ of hydrated and solvated proteins has impact on the interpretation of the dynamics. In particular, it has altered the assignment of the dynamic transition temperature $T_d$ in some cases [37].

The opportunity of uncovering the new break in some solvated proteins [37] is conferred by the fact that $T_g$ of the solvated protein is much higher than $T_{on}$, the onset temperature due to methyl group rotation, and yet significantly lower than $T_d$. The large gap in between $T_{on}$ and $T_d$ makes possible the unequivocal identification of the new break at $T_g$. The solvents achieving this feat include pure glycerol, and pure glucose, and aqueous mixtures of glycerol, glucose, or dissaccharides such as trehalose and sucrose [37]. However, this cannot be achieved when the solvent is pure $H_2O$ or $D_2O$, mainly because $T_g$ of the protein at full hydration usually falls below 200 K and is too close to $T_{on}$ to allow the break at $T_g$ to be identified unambiguously. For example, $T_g$ is 170 K for myoglobin at $h=0.40$ [33,42] and 165 K for myoglobin at $h=1$ [43,44]. $T_g$ is 169 K for hydrated methemoglobin (MetHb) with $h=0.50$ by Sartor et al. [30,44,45], and 172 K of hydrated metmyoglobin (MetMb) at $h=0.73$ by DSC [44]. $T_g$ is 150 K for lysozyme with more than 24 wt % of water by adiabatic calorimetry [31], 173 K by Brillouin scattering for lysozyme at $h=0.4$ [46] and 162 K for lysozyme at $h=0.64$ by DSC [44]. $T_g$ is ~ 200 K for bovine serum albumin (BSA) at $h=0.20$ by adiabatic calorimetry [32] and dielectric spectroscopy [48] and 175 K (onset) - 190 K (middle point) for BSA at $h=2-4$ by DSC [48,49]. $T_g$ is 170 K for Cytochrome C at 50% wt. hydration [43], and 186 K for fully hydrated ovalbumin [49].

The lack of clear evidence of the break of $<u^2(T)>$ at $T_g$ in the archetypal hydrated proteins is less than desirable in the quest to show that the phenomenon is general. In following up the previous study [37], the purpose of the present paper is to reexamine some outstanding neutron scattering data of hydrated proteins published with details that enable us to demonstrate the presence of the break of $<u^2(T)>$ at $T_g$ in addition to the dynamic transition at a higher temperature $T_d$.

The first set of data is on the dynamics of Ribonuclease A (RNase A) in dry and hydrated powder forms [50] measured by two spectrometers, IN16 and IN5, which have a 100-fold difference in accessible time-scales. By comparing the IN16 data of hydrated and dry RNase A, and reexamining the IN5 data of dry RNase A, the presence of the break of $<u^2(T)>$ at $T_g$ is deduced. The second set of data consists of the classical measurements of hydrated myoglobin at $h=0.4$ by Doster et al. [4,24] on IN13 and IN6 with accessible time-scales that differ by about 10-fold. In the same spirit as the RNase A data set, the absence of the methyl group rotation contribution to $<u^2(T)>$ in the data taken by IN6 makes it possible to see the break of $<u^2(T)>$ at $T_g$ of the hydrated myoglobin.

The third data set is taken from incoherent neutron scattering experiments on fully deuterated maltose binding protein (MBP) powder hydrated by $H_2O$, which primarily reflects the hydration water dynamics, and on a sample of hydrogenated MBP in $D_2O$ [13] Previously Wood et al. used the data from the two samples to show dynamic transition of the protein and its hydration water occurs at the same temperature $T_d$~220 K. Here we exploit feature of the $<u^2(T)>$ of the deuterated MBP and hydrated by $H_2O$ to show the presence of the break at $T_g$ ~200 K. The same two samples, H-MBP-$D_2O$ and D-MBP-$H_2O$, also had been measured on IN5 by Paciaroni et al.[11]. Like RNase A, the IN5 data help us to show unequivocally the presence of the break of $<u^2(T)>$ at the same $T_g$ ~200 K and the dynamic transition at a higher $T_d$~250-260 K.

The fourth set of data is on a completely deuterated purple membrane (PM) and hydrated by $H_2O$. PM is composed of a single membrane protein, bacteriorhodopsin (BR), and various lipid species. The elastic incoherent neutron scattering (EINS) experiments, performed on the IN16 spectrometer probe the dynamics of the hydration water [29,51].

The fifth and final set of data of a non-methyl-containing side chain, lysine, in bacteriorhodopsin (BR) by studying a completely deuterated PM in which the lysine residues are hydrogenated and comparing these dynamics to the dynamics of a natural abundance control PM sample [26].

**1.  IN16 and IN5 data of hydrated and dry RNase A**

Wood et al. [50] have made a comprehensive study of the dynamics of dry and hydrated RNase A in the full range of time and length-scales accessible by neutron spectroscopy. They achieved this goal by making measurements using a comprehensive set time-of-flight, backscattering and spin-echo spectrometers. For our present purpose, only their data of dry RNase A and RNase A hydrated by $D_2O$ with $h$=0.40 and measured on IN16 and IN5 are reexamined. IN16 and IN5 have respectively energy resolution of 0.9 and 100 μeV and accessible time-scales of 10 ps-1 ns and 0.1-6 ps.

We begin by reexamining the data of MSDs of dry RNase A and hydrated RNase A by $D_2O$ with $h$=0.40 from the IN16 backscattering spectrometer, which is reproduced in Fig. 1. It shows that at low-temperature, the $<u^2(T)>$ of the hydrated and dry samples are similar. As pointed out by Wood et al., fit by a straight line to the $<u^2(T)>$ between 20 and 70 K starts to deviate from the data of the dry sample at about 100 K, and from that of the hydrated sample at 125 K. This break at these low temperatures observed in both the dry and hydrated RNase A was attributed to methyl group rotation. The break is not well defined and less sharp than found in hydrated myoglobin [52]. This was rationalized by the methyl groups contribute only 18% of the scattering of RNase A, that is less than 25% in the case of myoglobin. The onset of the methyl group rotation at $T_{on}$~125 K for the hydrated protein gives rise to the stronger $T$-dependence of $<u^2(T)>$ past 125 K, which is suggested by the red line drawn starting from 125 K and ends at about 190 K. We observe $<u^2(T)>$ of the hydrated RNase A exhibits another break at some temperature between 175 and 180 K, as evidenced by a more rapid increase starting from there, which does not occur for $<u^2(T)>$ of the dry RNase A. This is suggested by the red line in the middle starting at 175 K and end at 225 K. The final break of $<u^2(T)>$ at about 220 K suggested by the intersection of the lines in the middle and on the right is the dynamic transition of the hydrated RNase A with $T_d$~220 K. The measured $<u^2(T)>$ of dry RNase A can be considered as contributed totally by methyl group contribution starting from 100 K and continue up to some higher temperatures past 180 K because the dynamic transition is suppressed in dry RNase A.

From this and the similarity of the $<u^2(T)>$ contributed by the methyl group rotation in the hydrated and dry RNase A up to 180 K, we assert that the break at a temperature between 175 and 180 is the change of $T$-dependence of $<u^2(T)>$ at $T_g$ of the hydrated RNase A that we are looking for.

Next we reexamine the $<u^2(T)>$ of the dry and hydrated RNase A obtained from the time-of-flight spectrometer IN5. The data from Wood et al. are reproduced in Fig. 2. Included in the figure are the $<u^2(T)>$ data from IN16 in Fig.1 shown again in Fig.2 (originally by Wood et al.) for comparison. Wood et al. pointed out that the analysis leading to $<u^2(T)>$ were performed in the same $Q$-range, and hence on the same length-scale, for both spectrometers. The differences in the $<u^2(T)>$ in Fig.2 arise solely from the large difference in energy resolutions and 100-fold difference in accessible time-scales of the two spectrometers. At low temperature, $<u^2(T)>$ from both spectrometers follow the same linear trend as a function of temperature up to approximately 100 K, reflecting contribution from protein atoms vibrating in harmonic potentials.

An important difference on the observation of the methyl group rotation by the two spectrometers is pointed out by Wood et al. [50]. The Arrhenius temperature dependence of the frequency of methyl rotations in amino acids has been obtained by NMR spectroscopy [53]. Using the Arrhenius dependence and the onset of methyl group rotation observed on the IN16 spectrometer with 1 μeV resolution is at $T_{on}$=100 K, Wood et al. deduced for the IN5 spectrometer with 100 μeV resolution that the onset temperature is much higher and approximately $T_{on}$≈250 K. From this, on IN5 they expect to see a break in the 250 K region for the dry sample. However, $<u^2(T)>$ of the dry sample in Fig. 2 shows no such clear break, and this discrepancy has puzzled them. Quite possibly, the observed $<u^2(T)>$ of the dry sample in the 250 K region do not come solely from the methyl group rotation. We suggest that, in analogy to what happens in glass-formers [54], a significant contribution to $<u^2(T)>$ comes from motions of molecules in the protein mutually caged by the anharmonic potential of their neighbors. The amplitude of this kind of motions increases monotonically with temperature [38] and may dominate the methyl contribution at 250 K, making the expected break unobservable.

More relevant to the present work is the $<u^2(T)>$ of the hydrated RNase A measured on IN5 in Fig.2. If one examine the temperature variation of $<u^2(T)>$ closely, two breaks can be discerned. The three red lines are used to bring out the breaks. The break at higher temperature is the dynamic transition observed on IN5 with $T_d$≈260 K, and is the correspondent of the dynamic transition on IN16 with $T_d$≈225 K. Nearly the same $T_d$≈260 K was found by Tsai et al. [55] in RNase A hydrated with $D_2O$ to 24 wt % on the Fermi Chopper spectrometer at the NIST Center for Neutron Research with energy resolution of 140 μeV comparable to 100 μeV of IN5. The break in the $<u^2(T)>$ of the hydrated RNase A by Tsai et al. also seems to be present at $T_g$~175-180 K, although less certain. According to our interpretation [56,57], the dynamic transition of hydrated proteins is caused by the secondary relaxation of the hydration water entering into the time window of the spectrometer. On changing from IN16 to IN5 and reducing the time-scale by 100-fold, a higher temperature is required for the β-relaxation time $\tau_\beta$ of hydration water to enter the time window of IN5, and hence the shift of $T_d$ from 225 to 260 K. The two values of $T_d$ are consistent with the temperatures at which $\tau_\beta$ has reached 1 ns and 6 ps within an order of magnitude respectively [57].

Our focus is on the break at lower temperature of about 180 K in Fig.2, which turns out to be about the same as the temperature of the break of the same hydrated sample we have identified before from $<u^2(T)>$ on IN16 in Fig.1. This is not an accident if both breaks are at $T_g$ of the hydrated RNase A because this break is always near $T_g$, independent of the energy resolution of the spectrometer [38]. Furthermore, we are guaranteed by the analysis of the dry sample that the methyl group rotation does not contribute to $<u^2(T)>$ on IN5 except beyond about 250 K. Thus, the break at 180 K of $<u^2(T)>$ of the hydrated RNase A on IN5 cannot be anything else but at $T_g$. Surprisingly we cannot find any report on calorimetric or dielectric measurement of the glass transition temperature of hydrated RNase A by $H_2O$ or $D_2O$ with $h$=0.40. Such measurements if carried out will additionally confirm our identification of the new break is in fact at $T_g$ of the hydrated RNase A. However a careful X-ray crystallography investigation for RNase A, solvated by (v/v) 2-methyl-2,4-pentanediol-water at 50%, by Tilton et al. [58] emphasized the existence of a biphasic behavior in the Debye-Waller factor as a function of temperature. They found the protein molecule expands slightly and linearly with increasing temperature due primarily to subtle repacking of the molecules, and a break occurs at a characteristic temperature of 180-200 K. Since crystallographic measurements are mainly sensitive to structural changes, we may take this temperature as a reliable estimate of $T_g$ near 180 K for hydrated RNase A.

High frequency dielectric measurements of the β-relaxation time $\tau_\beta$ of hydration water of RNase A will help to explain the dynamic transition occurring near 225 and 260 K found on IN16 and IN5 respectively.

## 2. IN13 and IN6 data of hydrated myoglobin

There is another past experimental situation similar to that discussed in the previous section in which the contribution to $<u^2(T)>$ from methyl group rotation is moved to higher temperatures and suppressed. This is the investigation of myoglobin hydrated with $H_2O$ and $D_2O$ (0.35 g/g) by using the low resolution, high-flux time-of-flight (TOF) spectrometer (IN6) accessing times faster than about 15 ps. The 15 ps TOF displacements of $H_2O$ and $D_2O$ hydrated myoglobin published in Ref. [24] and reproduced in Fig.3 show $<u^2(T)>$ in both systems increase in parallel with a common onset at ~180 K. The common temperature dependence is indication of the coupled dynamics of the protein with that of the hydration water found in other solvated proteins. As pointed out by Doster, this transition is unrelated to methyl group dynamics because the correlation time for methyl group rotation at ~180 K is of the order of 100 ps [59], which is outside the IN6 spectral window. The onset at 180 K is accompanied by the increase in intensity of the high frequency spectrum suggesting a fast process is responsible for it.

The $<u^2(T)>$ of the same two samples hydrated with $H_2O$ and $D_2O$ hydration had been measured also by IN13. A procedure was used to subtract off the contribution of the methyl groups from the measurements, and the resulting $<u^2(T)>$ also are shown in Fig.3. Again both $<u^2(T)>$ of protein and water displacements are similar in magnitude and exhibit the same temperature dependence. The dynamic transition temperature of the same hydrated myoglobin becomes 240 K, comparable to that observed in other fully hydrated proteins by IN13. The deduced value of $T_d$ from IN13 is higher than that from IN6, despite the former has a longer observation time. This contradicts the common observation in hydrated and solvated proteins that $T_d$ is shifted to lower temperature when changing to a spectrometer which has longer observation times[14,18,22,60]. To resolve the conundrum he [24] found by comparing IN6 and IN13 results in hydrated myoglobin, Doster suggested that in experiments the time-of-flight IN6 is sensitive to much faster motions than the back-scattering IN13. The faster motions he had in mind are H-bond fluctuations, which he considered to be closely related to reorientation and diffusion of water near the protein surface. Using the super-Arrhenius dependence for its relaxation time determined by 2H-NMR, and assuming displacements are resolution-limited, he was able to obtain nearly perfect fit to the data as shown by the solid line in Fig.3. Actually, this procedure seems to be rather questionable, as the elastic intensity measured on IN13 is as sensitive to faster motions as the one from IN6, and for this reason the puzzle remains. We suggest that the conundrum of the observation of the onset at ~180 K by IN6 can be resolved by the presence of the change of $T$-dependence of $<u^2(T)>$ at $T_g$ in $H_2O$- and $D_2O$ hydrated myoglobin (0.35 g/g). Such change of $T$-dependence of $<u^2(T)>$ at $T_g$ has been found also in conventional glassformers by IN6 as well as in IN13 [38]. Moreover, glass-liquid transition of 170 K in hydrated myoglobin powder (0.4 g/g) and myoglobin crystals was found by DSC by Doster and coworkers [33,42], and at 169 K of hydrated methemoglobin (MetHb) with 0.50 h, by Sartor et al. [30,45]. The data of the specific heat $\Delta C_p$ from DSC experiment by Doster and coworkers of adsorbed water in hydrated myoglobin powder (0.4 g/g) and myoglobin crystals from calibrated differential calorimetry experiments are reproduced in the inset of Fig.3. The value of 170 K for $T_g$ determined by them is indicated by the arrow [24]. Thus, as suggested by the arrow located near 180 K in Fig.3, the rise of $<u^2(T)>$ seen starting there can be the change of $T$-dependence of $<u^2(T)>$ near $T_g$ that we are looking for. On the other hand, we can estimate by the JG β-relaxation time of water given in Ref. [57], that $T_d$ on IN6 is near or higher than 260 K, *i.e.* higher than the 240 K found on IN13, as suggested by the red arrow in Fig.3. Such value of ~260 K for $T_d$ for hydrated myoglobin on IN6 is supported by observation of dynamic transition in other hydrated proteins on IN6 at comparable temperatures. An example is $T_d \approx 260$ K for heparan sulfate (HS) hydrated to 0.43 g $D_2O$/g HS (HS-0.43 g/g) [22]. By using IN13 instead of IN6, $T_d$ of the hydrated heparan sulfate is down shifted to 240 K [22,60], which is in accord with $T_d$=240 K of $H_2O$ and $D_2O$ hydrated myoglobin (0.35 g/g) from IN13 by Doster in Fig.3.

   The reason for Doster [24] to remove the contribution of methyl groups from the IN13 data for comparison with the IN6 data in Fig.3 is because the latter has no contribution from the methyl group rotation. To do this, he assumed the $<u^2(T)>$ obtained also by IN13 of dry and vitrified myoglobin is entirely coming from methyl rotations, and he used it to subtract off from the $<u^2(T)>$ of the hydrated myoglobin. The resulting $<u^2(T)>_{DT}$, which should bring information only about the dynamic transition on IN13, is shown in Fig.3. It can be sen by inspection that $<u^2(T)>_{DT}$ is smaller than $<u^2(T)>$ from IN6, which is a surprise. Actually, the procedure he used is based on two major assumptions. As we mentioned above, the first one is that the entire $<u^2(T)>$ up to 320 K in dry and vitrified myoglobin is due to methyl rotation (except for the vibrational contribution). $T_g$ is higher in dry and vitrified myoglobin, and this can lead to continued rise of $<u^2(T)>_{NCL}$ contributed from the nearly constant loss (NCL) of caged dynamics [62] past $T_g$ of the hydrated myoglobin near 180 K. Such approximately linear increase of $<u^2(T)>_{NCL}$ with $T$, which is found in conventional glassformers [38], is also evident from molecular dynamics simulations of hydrated proteins [13] discussed in our previous paper [37] and also in purple membrane [51] to be discussed later. Thus, an overestimate of the methyl rotation contribution could have been made on assuming it is the entire $<u^2(T)>$ up to 320 K in dry and vitrified myoglobin. Another assumption is that the contributions to mean square displacement from the methyl rotation is additive to the contribution from NCL, which causes the transition at $T_g$, and also to that from the relaxation process causing the dynamic transition at $T_d$. This may not be true, as it would correspond to a perfect decoupling between methyl group motions and the other relaxation processes. Adopting both the assumptions in analyzing data, as Doster did, the possible change of the $<u^2(T)>$ temperature-dependence at $T_g$~180 K may be obscured.

3. **IN16 data of fully deuterated MBP and hydrated by $H_2O$**

Water and protein motions in hydrated maltose binding protein (MBP) have been probed separately by measuring fully deuterated MBP sample hydrated in $H_2O$ and natural-abundance MBP sample hydrated in $D_2O$, respectively by Wood et al.[13]. The $<u^2(T)>$ data on IN16 of the two samples were used by Wood et al. to show that the dynamic transition of the protein and its hydration water occurs at the same temperature $T_d$~220 K, thus supporting strong coupling between protein and solvent as far as the dynamics transition is concerned. In Figure 4 we reproduce their data of the mean square displacements in maltose binding protein from the H-MBP-$D_2O$ sample, and in its hydration water from the D-MBP-$H_2O$ sample.

Now we closely examine the $<u^2(T)>$ of the fully deuterated MBP and hydrated by H$_2$O to show the presence of the break at $T_g$. The $<u^2(T)>$ of the fully deuterated MBP sample hydrated in H$_2$O have no contribution from the methyl group rotation, and hence it offers the possibility of finding the break at a temperature below $T_d$~220 K. If found, the temperature at the break is identifiable with $T_g$. Naturally $T_d$ is higher than $T_g$ because the relaxation time of the hydration water at $T_g$ is much longer than the time-scale of 1 ns of IN16.[56,57] Only at the higher temperature, $T_d$~220 K, the spectral contribution from hydration water relaxation starts entering the time window of the IN16 spectrometer to give rise to the dynamic transition. The green line in Fig.4 depicts the possible harmonic contribution to $<u^2(T)>$ of the hydration water. It does represent the data sufficiently well up to about 190 K where $<u^2(T)>$ starts to deviate and the break at the 190-200 K region becomes evident. The milder increase of $<u^2(T)>$ after this first break eventually gives way to a stronger increase with temperature at about 220-225 K, which corresponds to the dynamic transition identified by Wood et al. The possibility that another change of $<u^2(T)>$ have taken place at temperatures lower than $T_d$~220 K may have already been recognized by Wood et al. This can be drawn from their statement: "MBP exhibits a dynamical transition in the same temperature range (180-220 K) as other soluble proteins". We suggest that this broad transition is actually composed of the break at $T_g$ and the dynamic transition at $T_d$. At 190 K, for water in various aqueous mixtures and hydrated proteins as well as confined in nano-meter spaces, its shortest relaxation time ever recorded is about $10^{-6}$ s [57]. It is three decades longer than 1 ns and likely it cannot be responsible for the break of $<u^2(T)>$ starting at 180-190 K seen by IN16.

It is interesting to observe the difference in the temperature dependence between $<u^2(T)>$ of the natural-abundance MBP sample hydrated in D$_2$O and $<u^2(T)>$ of the fully deuterated MBP and hydrated by H$_2$O in the temperature range 150-200 K. As shown in Fig.4, $<u^2(T)>$ of the former exhibits a break due to the methyl group rotation at approximately 150 K (absent in the latter) and it continues to rise smoothly over and above $<u^2(T)>$ of the latter, making the observation of the break at $T_g$ in the range 190-200 K impossible.

At present there is neither published calorimetry nor dielectric relaxation data of hydrated MBP to independently determine $T_g$. A worthwhile undertaking is to make DSC and dielectric relaxation measurements on hydrated MBP to check if $T_g$ is in the neighborhood of 190-200 K, and the secondary relaxation time of water is near 1 ns at $T_d$~220-230 K. Notwithstanding, support can be found from the molecular dynamics simulation of MBP hydrated to about one hydration layer per MBP molecule also by Wood et al. [13]. These authors presented the time evolution of mean-square displacements of water O atoms over a range of temperature from 150 to 300 K. The authors reported that 240 K is the temperature $T_d$ of the dynamic transition in the simulations for times up to 100 ps. These times for simulation are comparable with the time window of the IN13 spectrometer, using which $T_d$=240 K is often observed on fully hydrated proteins. The temperature dependence of the water mean square displacements of the O atoms of the water molecules at 100 ps is also presented by them. In addition to the dynamic transition at $T_d$=240 K, the mean square displacements show a change of $T$-dependence at about 190-200 K in the simulation. As discussed in more details in the preceding paper [37], this finding by simulations supports the presence of the change of $T$-dependence of $<u^2(T)>$ at $T_g$, and $T_g$ is in the neighbourhood of 190-200 K.

The same two samples, H-MBP-D$_2$O and D-MBP-H$_2$O, also had been measured on IN5 by Paciaroni et al. [11]. According to these authors, contributions to the measured elastic intensities, and calculated mean square displacement of MBP, $<u^2>_{MBP-IN5}$, and its hydration water, $<u^2>_{H2O-IN5}$, only can come from motions faster than about 15 ps. As discussed before in Section 2 on RNase A, the protein elastic intensity and $<u^2>_{MBP-IN5}$ measured on IN5 has no contribution from the methyl group rotations at temperatures lower than 250 K. This creates a favorable condition for observing on IN5 the break of $<u^2>_{MBP-IN5}$ at $T_g \approx 200$ K in both H-MBP-D$_2$O and D-MBP-H$_2$O, which indeed was realized and is now shown in Fig.5. Presented in this figure are the temperature dependence of the elastic intensities, integrated over a range of rather small Q values (0.4 Å$^{-1}$ < Q < 1.0 Å$^{-1}$), for both H-MBP+D$_2$O and D-MBP-H$_2$O (*i.e.*, the protein hydration water). There is almost linear T-dependence of the intensity at low temperatures, but deviations start noticeably at ~200 K, and increase with temperature. Two breaks in the T-dependence of the intensity of both MBP(H)+D$_2$O and the protein hydration water can be discerned, as suggested by the lines drawn in Fig.5. One occurs near 200 K, which we identify once more as due to the change in T-dependence at $T_g$ of the nearly constant loss (NCL) of caged dynamics of the coupled protein and hydration water. Quite interestingly, the break is seen at approximately the same temperature ~200 K on IN5 and IN16 (see Fig.4), thus confirming the general property [38] that this change in the T-dependence is invariably near $T_g$, independent of the energy resolution of the spectrometer used. On the other hand, the second break in the IN5 intensity of both H-MBP+D$_2$O and D-MBP-H$_2$O at just above 250 K is the dynamic transition. The shift of $T_d$ from ~220 K on IN16 to ~250 K on IN5 clearly demonstrates the dependence of $T_d$ on the observation time-scale of the dynamic transition. This is also consistent with the dispersion of the JG β-relaxation of the hydration water starts entering the time window of 1 ns on IN16 at ~220 K and 15 ps at a higher temperature of ≥ 250 K on IN5 [62].

The inset of Fig.5 shows the approximately linear temperature dependence of $<u^2>_{MBP-IN5}$, and $<u^2>_{H2O-IN5}$ at lower temperatures, and departures at temperatures above 200–220 K. Reproduced also is the temperature dependence of $<u^2>_{MBP-IN5}$, and the mean square displacement of MBP in H-MBP-D$_2$O, $<u^2>_{MBP-IN16}$, measured instead on the neutron backscattering spectrometer IN16 by Wood et al., and presented before in Fig.4. The measured $<u^2>_{MBP-IN16}$ by IN16 can detect motions faster than about 1 ns. It can be seen that starting from about 150 K, the protein $<u^2>_{MBP-IN16}$ measured on IN16 becomes increasingly larger than the protein $<u^2>_{MBP-IN5}$ measured on IN5, indicating contribution from relaxation with characteristic times longer than 15 ps but shorter than 1.5 ns. As discussed above on hydrated RNase A and myoglobin, the relaxation corresponding to the methyl group reorientations is located in some temperature range above ~150 K. Despite the trend of $<u^2>_{MBP-IN5}$ is much more scattered than that of the elastic intensity, both the departure at $T_g \approx 200$ K and the dynamic transition at $T_d \geq 250$ K can be reasonably well revealed. The lines drawn in the inset serve the purpose of delineating the two breaks in $<u^2>_{MBP-IN5}$.

The advance made here is the recognition of the presence of the break in both the elastic intensity and $<u^2>_{MBP-IN5}$ measured on IN5 at $T_g \approx 200$ K, and the support by the analogy to the data of RNase A on IN5 shown before in Fig.2. [50] With this recognition, we can avert the equivocal conclusion of Ref. [11]: "Our results show that the dynamics of both protein and hydration water undergoes a transition in the same temperature range, in agreement with the results obtained by Wood et al.".

### 4. IN16 data of completely deuterated purple membrane hydrated by H$_2$O

Dynamics of the protein and hydration water in purple membrane (PM) have been a subject of intense study by neutron scattering [26,29,51,63,64]. PM is composed of a single membrane protein, bacteriorhodopsin (BR), and various lipid species. Here we focus our attention on the sample of completely deuterated purple membrane (PM) and hydrated by H$_2$O denoted by D-PM-H$_2$O, and the $<u^2(T)>$ of the hydration water from the elastic incoherent neutron scattering (EINS) experiments performed on IN16 [26,29,51]. The data of interest are reproduced in Fig.6 and are illustrated with the assist of lines to show the presence of two breaks. The break at higher temperature near 250-260 K appears not only in the $<u^2(T)>$ of hydration water in D-PM-H$_2$O but also in H-PM-D$_2$O, an identically prepared control sample of natural-abundance PM hydrated in D$_2$O. Both samples are hydrated to approximately 0.3 g of water per g of PM. This is the dynamic transition of the hydrated PM, showing up at nearly the same temperature $T_d \approx 260$ K in the two samples, and caused by the coupling between hydration water and protein[29,51]. We identify the break of $<u^2(T)>$ of hydration water in D-PM-H$_2$O at the lower temperature near 200 K as the one at $T_g$ that we are looking for. The opportunity to observe this in D-PM-H$_2$O is due to the absence of the contribution from methyl group rotation by complete deuteration of PM. On the other hand, the contribution from methyl group rotation to $<u^2(T)>$ in H-PM-D$_2$O clearly shows up in Fig.6 by the strong increase with temperature starting with the break at 120 K. The size of the increase of the contribution to $<u^2(T)>$ from the methyl group rotation is large compared with that from the caged dynamics, which can be estimated from the $<u^2(T)>$ of hydration water in D-PM-H$_2$O in Fig.6. Thus the break of the contribution to $<u^2(T)>$ from caged dynamics at $T_g$ cannot be resolved from $<u^2(T)>$ in H-PM-D$_2$O. We think this is the reason for not observing the break or inflection of $<u^2(T)>$ in H-PM-D$_2$O at $T_g \approx 200$ K, and hence there is perhaps no need for Wood et al. to make the PM an exception by concluding that "…water and membrane motions on the ps–ns time scale are not directly coupled to each other and that the latter are thus neither solvent-slaved, nor hydration-shell coupled at temperatures <260 K.". Molecular dynamics simulation of the mean square displacements (MSDs) of the centers-of-mass of water molecules in the first hydration shell of PM was also presented by Wood et al.[51]. The MSDs at $t =30$ ps when plotted against temperature exhibit a break at 200 K. The chosen time of 30 ps is much shorter than the longest time of 1 ns accessible to IN16. The fact that the same temperature of 200 K for the break found by EINS on IN16 and by molecular dynamics simulation is interesting but unsurprising. This is because, as found before in all kinds of glassformers [38] and illustrated for hydrated proteins in the previous sections, also in the case of hydrated PM the break at $T_g$ is independent of the energy resolution of the spectrometer.

    Direct experimental evidence that $T_g$ of the hydrated PM studied by Wood et al. is about 200 K can be drawn from the study of PM hydrated to $h$=0.4 and 0.2g H$_2$O/g of PM by dielectric spectroscopy and differential scanning calorimetry between 120 and 300 K by Berntsen et al. [65]. Discussed before in our preceding paper [37], they found by calorimetry a pronounced endothermic process at 190-200 K, which can be identified as $T_g$ of the hydrated PM. By dielectric spectroscopy they found the JG β-relaxation of the hydration water with relaxation time $\tau_\beta$ showing Arrhenius $T$-dependence at low temperatures with activation energy of about 54 kJ/mol, typical of this kind of relaxation, as well as changing to a stronger $T$-dependence at high temperatures after crossing 190–200 K [57,57,62,66]. This behavior is another indication that $T_g$ is within the range of 190-200 K because change of $T$-dependence of $\tau_\beta$ at $T_g$ is a universal feature found in all glassformers including aqueous mixtures and hydrated proteins.

    Since two breaks at ~200 K and ~250-260 K have been found in the $<u^2(T)>$ of hydration water in D-PM-H$_2$O by EINS, it would be helpful to have independent experimental data to distinguish the characters of the two breaks. Specific question to ask is which break is the dynamic transition at $T_d$? Although from the relation, $T_g< T_d$ [57], it is likely that the break at 250-260 K defines $T_d$. Independent confirmation comes from Mössbauer spectroscopy experiments on PM [67]. The experiment studied the dynamics of one kind of iron bound to BR either to the acid side chains and/or the phosphate group of the lipids, and another kind linked non-specifically to the carboxy groups of the protein. The experiments showed dynamical transitions of BR at some temperature $T_d$ within the range, 200-220 K,

depending on the oxidation state of the Fe ion. From the 140 ns time-scale of Mössbauer spectroscopy experiments much longer than the ~1.5 ns of EINS experiment on IN16, $T_d$ is expected to be significantly lower by the former than the latter, according to the explanation of the dynamic transition by the JG β-relaxation of hydration water entering the time window of the spectrometer used [56,57,62]. On examining the dielectric relaxation data by Berntsen et al. [65], the JG β-relaxation times of hydrated PM at $h$=0.2 and 0.4 are indeed close to 140 ns at 200-220 K. Thus, the Mössbauer spectroscopy experiments confirm that the break at ~250-260 K in the $<u^2(T)>$ of hydration water in D-PM-H$_2$O by EINS is the dynamic transition.

## 5. IN16 data of a completely deuterated PM with hydrogenated lysine residues hydrated by D$_2$O

In a study recently performed on IN16 the PM residues have been selectively deuterated to properly highlight their specific contribution [26]. In this system only leucine or isoleucine residues contain methyl side chains, the non-methyl-containing side chain being localized in bacteriorhodopsin (BR). The investigated sample (Lys-PM) was a D$_2$O hydrated perdeuterated PM, with hydrogenated lysine residues. Therefore, the measured $<u^2(T)>$ reflects principally the dynamics of the non-methyl-containing side chain, lysine. The results of $<u^2(T)>$ from lysine in Lys-PM are reproduced in Fig.7. It is compared with the $<u^2(T)>$ of a D$_2$O-hydrated natural abundance control PM sample, same as that presented before in Fig.6. Now the probed dynamics is the one of the non-methyl-containing side chain in lysine (sample Lys-PM), which is to be distinguished from the dynamics of hydration water probed in the sample D-PM-H$_2$O, already discussed in the previous section. Nevertheless, it is clear by inspection of Fig.7 that a break in the temperature dependence of $<u^2(T)>$ in lysine of Lys-PM appears near 200 K, in strong analogy with the $<u^2(T)>$ of hydration water in D-PM-H$_2$O shown in Fig.6. This remarkable resemblance provides evidence that the dynamics of the protein and the hydration water are coupled not only at $T_d$ but also down to the region around $T_g$.

## 6. Summary and Conclusion

In conventional glassformers, the dynamics of molecules while mutually caged by the intermolecular potentials show up isothermally in the mean square displacement $<u^2(t;T)>$ as function time or in the susceptibility $\chi''(\omega;T)$ as function of frequency as power laws given by $At^c$ and $B\omega^{-c}$ respectively with $0<c<<1$. This caged dynamics regime holds at times before the relaxation time $\tau_\beta$ of the Johari-Goldstein (JG) β-relaxation, which is the first genuine relaxation process that is effective in dissolving the cages. Despite the caged dynamics occurs at much shorter times than the structural α-relaxation time $\tau_\alpha$, it is a general experimental fact that its intensity senses the glass transition. Specifically, $<u^2(t;T)>$ or $\chi''(\omega;T)$ of caged dynamics changes temperature dependence when temperature is raised to cross the glass transition temperature $T_g$. This general property of caged dynamics is related to change of $T$-dependence of $\tau_\beta$ and the relaxation strength of the JG β-relaxation also generally found at $T_g$. Since hydrated and solvated proteins are glassformers, it is natural to inquire whether both the caged dynamics and the JG β-relaxation in them exist and exhibit the same changes on crossing $T_g$. If so, the similarity in the dynamics between solvated proteins and conventional glassformers can be brought closer together. This similarity to conventional glassformers has been realized for the JG β-relaxation of hydrated and solvated proteins. For the caged dynamics, the similarity is largely unexplored until recently when we have re-examined the incoherent neutron scattering data of solvated proteins [37]. Clear evidence of the change in temperature dependence of either the elastic scattering intensity or $<u^2>$ has been found in a number of solvated proteins. The solvents are either glycerol or monosaccharides and their mixtures with water, having $T_g$ of the solvated protein higher than when only water is used. Raising $T_g$ of the solvated protein is the key to the success of identifying the change of $T$-dependence of $<u^2>$ at $T_g$ in the forerunning study. Higher $T_g$ comes with certain advantages in detecting the change of $<u^2>$ at $T_g$. Firstly, the undesirable contribution to $<u^2>$ from rotation of the methyl groups in the protein at lower temperatures can be circumvented. Secondly, $T_d$ of the dynamic transition is also higher, more separated from $T_g$, and hence making the break in $T$-dependence of $<u^2>$ at $T_g$ to be distinguished from that at $T_d$. Thirdly, the change of $T$-dependence of $<u^2>$ at $T_g$ becomes more prominent at higher $T_g$. All these advantages disappear in proteins hydrated by pure water where often $T_g$ is lower. Consequently, clear evidence of change of $T$-dependence of $<u^2>$ at $T_g$ is hard to find from neutron scattering data of hydrated proteins. Nevertheless, all is not lost. By manipulating the sample such as by full and partial deuteration of the protein, or by using spectrometers sensitive to motions of shorter time-scales, the change of $T$-dependence of $<u^2>$ at $T_g$ has been successfully detected in several hydrated proteins. From the results we conclude that the phenomenon is present in hydrated and solvated proteins, although it may be difficult to be resolved in some cases. It originates from motion of caged molecules which has no characteristics time, unlike relaxation process. Nevertheless its amplitude of motion is sensitive to glass transition, and shows up by the change of $T$-dependence on crossing $T_g$, independent of the timescale of the spectrometer used. In contrast, the dynamic transition depends on the energy resolution of the spectrometer. This is because the onset of the dynamic transition is caused by entrance of the relaxation of the solvent coupled to the protein into the time window of the spectrometer. Since the relaxation time is thermally activated, shorter time window of the spectrometer requires higher temperature $T_d$ for the relaxation to enter it and trigger the dynamic transition. This will be further demonstrated quantitatively in a forthcoming paper for the cases considered here and in the forerunning paper[37]. Since the change of $T$-dependence of $<u^2>$ at $T_g$ is detected from either the protein or the solvent, another important conclusion is that the dynamics of the protein and the solvent are coupled not only in the dynamic transition at $T_d$ but also at $T_g$.

**Acknowledgment**

KLN thank CNR-IPCF at Pisa, Italy, and Prof. Pierangelo Rolla and Prof. Mauro Lucchesi of the Dipartimento di Fisica, Università di Pisa, Italy for hospitality during his stay in the period from March 9th-June 1st, 2011.

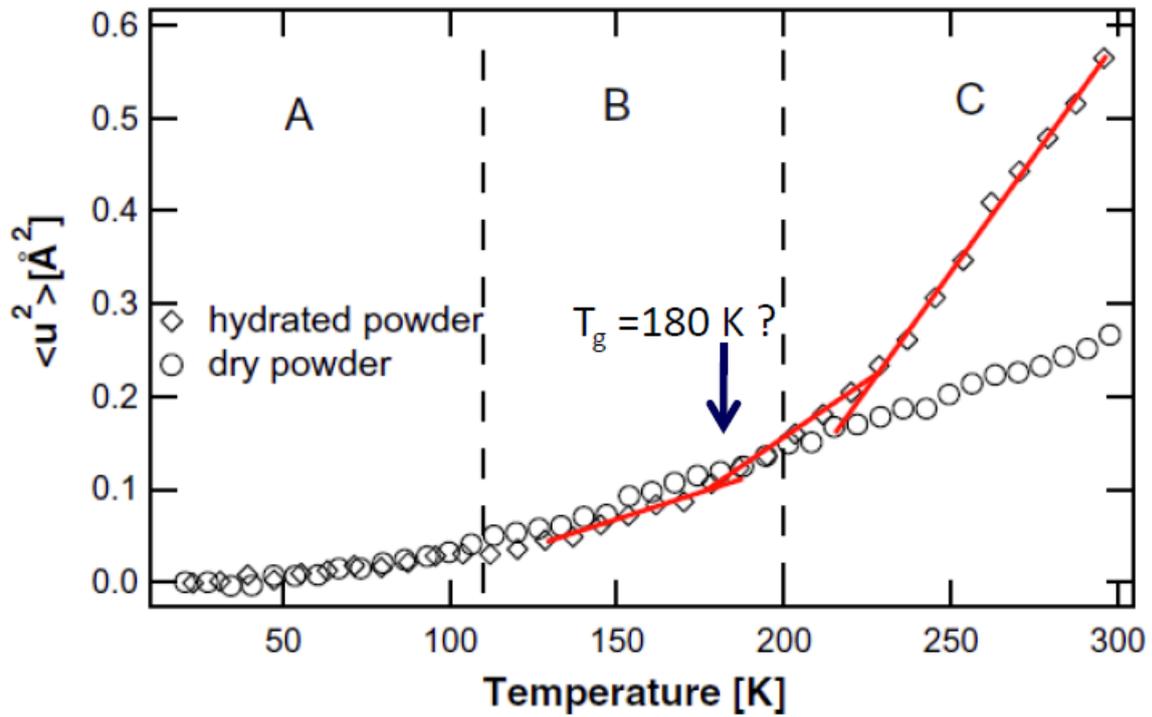

**Figure 1.** Mean square displacement as a function of temperature for dry (circles) and hydrated (diamonds) powders of RNase A measured on IN16. Reproduced from Ref.[51] by permission. The lines not in the original figure serve no other purpose except to suggest possibly the presence of break in $T$-dependence for the hydrated powders at $T_g \sim 180$ K.

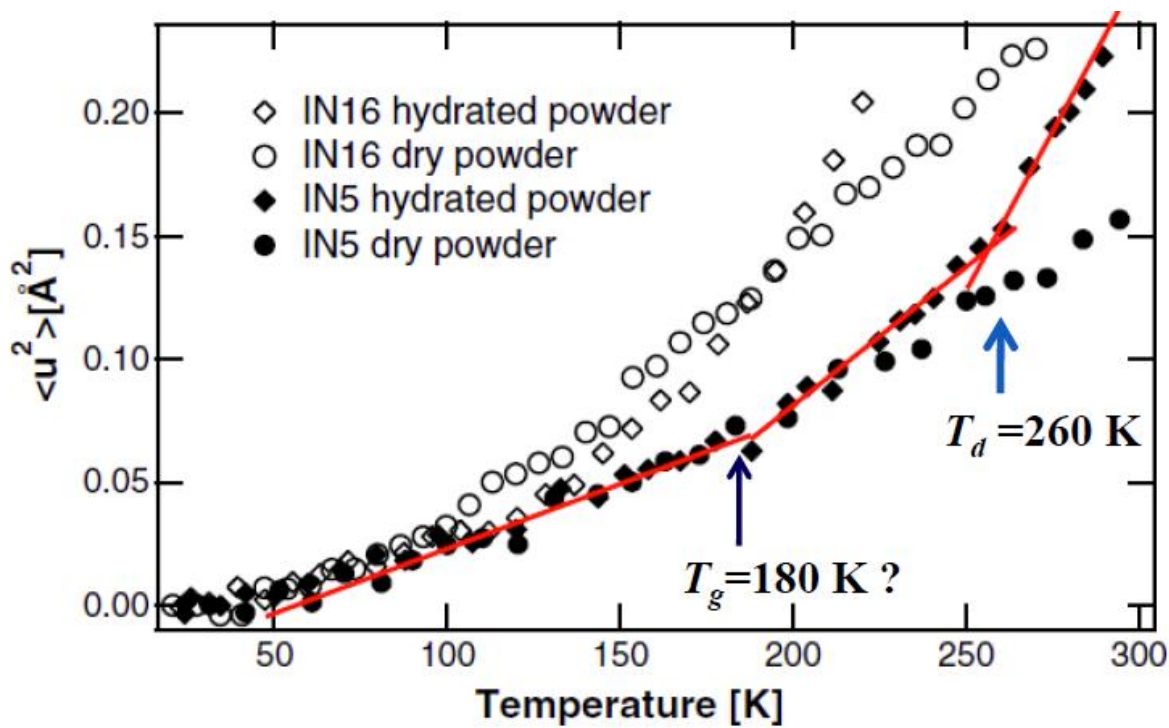

**Figure 2.** Mean square displacement extracted from the analysis of IN5 data for hydrated (full diamonds) and dry (full circles) powder RNase A samples. Data from IN16 are also shown for comparison with hydrated sample shown by empty diamonds, and the dry sample by empty circles. Reproduced from Ref.[51] by permission. The lines not in the original figures serve no other purpose except to suggest the presence of break in $T$-dependence for the deuterated RNase A at $T_g$~180 K, and the dynamic transition at $T_d$~260 K.

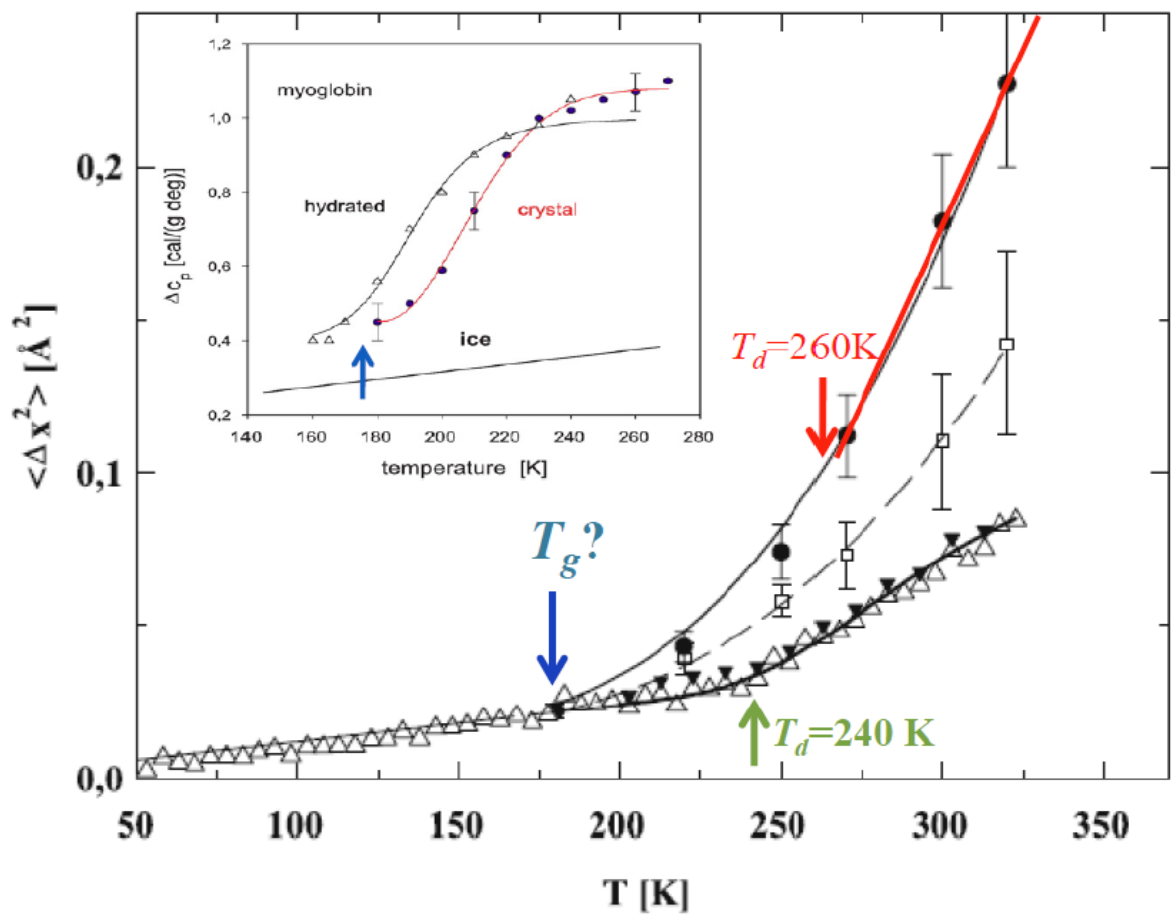

**Figure 3.** Mean square displacements in hydrated myoglobin (0.35 g/g). IN6: $D_2O$–hydrated myoglobin (open squares) and hydration water of $H_2O$–hydrated myoglobin (closed circles). IN 13: $D_2O$-(open triangles) and $H_2O$–hydrated (full triangles) myoglobin (0.35 g/g). The Gaussian component with contribution of methyl groups has been removed. The full lines represent the fits to an asymmetric two-state model (IN6) and to a water-coupled process, limited by the instrumental resolution (IN13). Inset: Specific heat of adsorbed water in hydrated myoglobin powder (0.4 g/g) (triangles) and myoglobin crystals (filled circles) derived from calibrated differential calorimetry experiments. The background of the dry protein was subtracted. Reproduced from Ref. [24] by permission. The red line and arrows not in the original figures serve no other purpose except to suggest the presence of break in $T$-dependence at $T_g$~180 K, and the dynamic transition at $T_d$~260 K for the data on IN6, and $T_d$~240 K for the data on IN13.

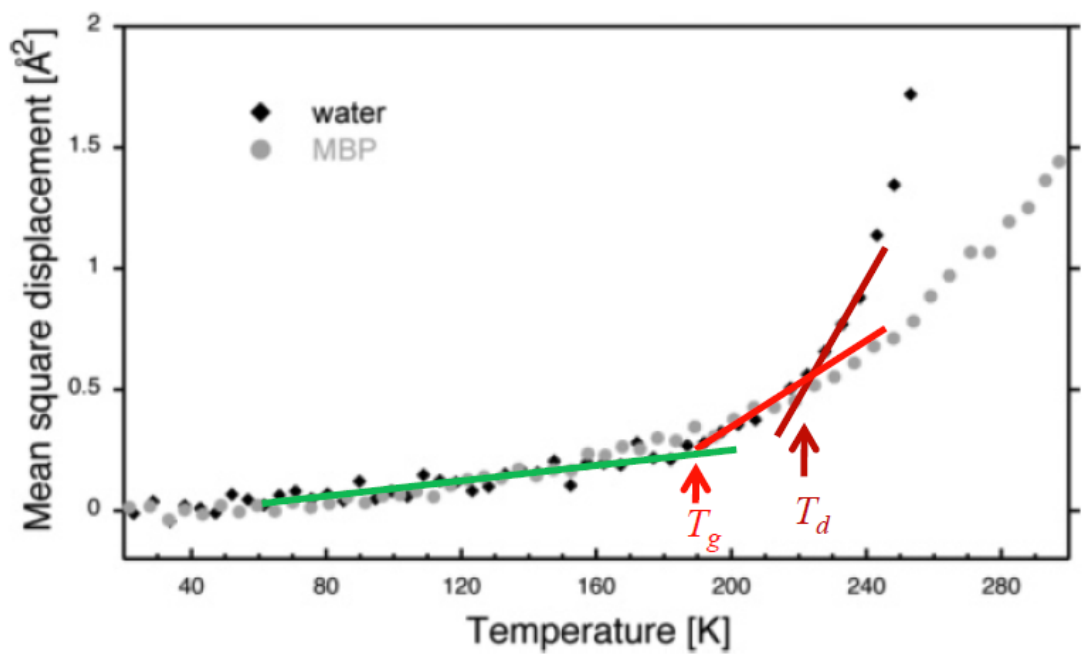

**Figure 4.** Mean square displacements of ns - ps motions in maltose binding protein (H-MBP-$D_2O$ sample gray circles) and in its hydration water (D-MBP-$H_2O$ sample; black diamonds). Reproduced from Ref.[13] by permission. The lines not in the original figures serve no other purpose except to suggest the presence of two changes in slope of the temperature-dependent mean square displacement of the hydration water in D-MBP-$H_2O$, one at $T_g$~200 K and another one at $T_d$~220 K is the dynamic transition identified before by Wood et al.[13].

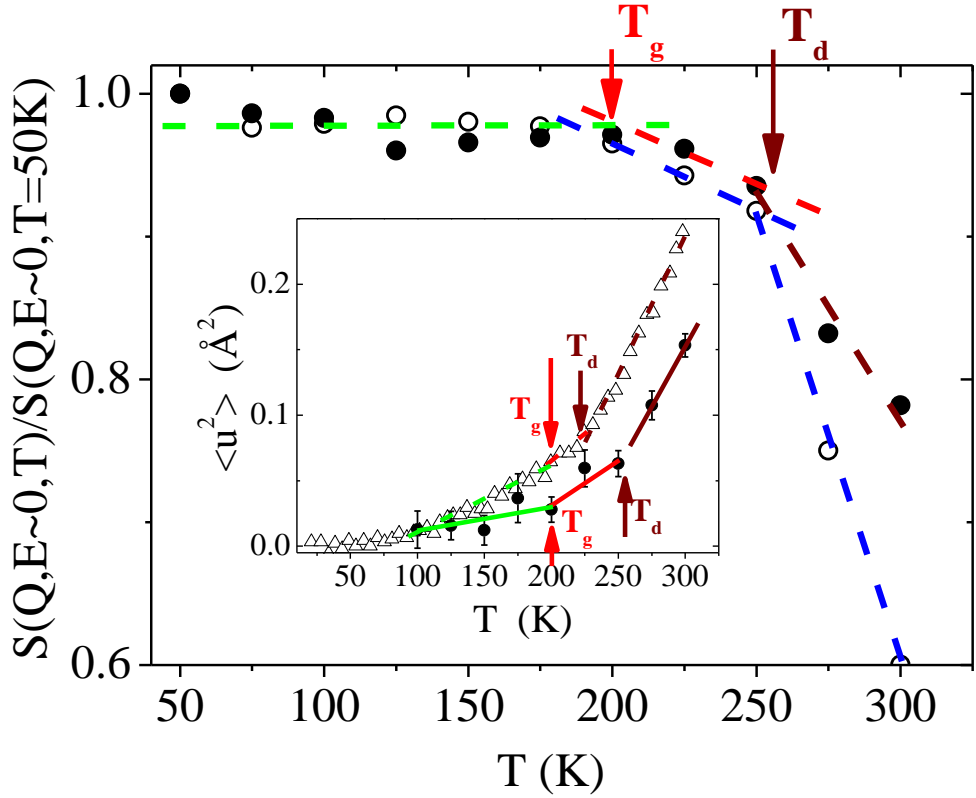

**Figure 5.** Incoherent elastic intensities measured on IN5, normalized with respect to the lowest temperature, integrated over a small $Q$-range (0.4 Å$^{-1}$ < $Q$ < 1.0 Å$^{-1}$), of MBP hydration water (empty circles) and H-MBP-D$_2$O (full circles). Inset: Mean square displacements versus $T$ of H-MBP-D$_2$O as calculated from measurements on IN5 (full circles) reported by Paciaroni et al. [11], and on IN16 (empty triangles) reported by Wood et al. [13]. The latter were rescaled by a factor of six to enable comparison with the former. For IN5 data in both the main figure and the inset, the lines serve no other purpose except to suggest the presence of two breaks in $T$-dependence of the MSD for H-MBP-D$_2$O, one at $T_g$~200 K and another one at $T_d$~250-260 K. For the IN16 data in the inset, the lines are used to suggest the presence of two breaks in $T$-dependence of the MSD for H-MBP-D$_2$O, one at the same $T_g$~200 K and another one at a lower $T_d$~220 K, shown before in Fig.4
17

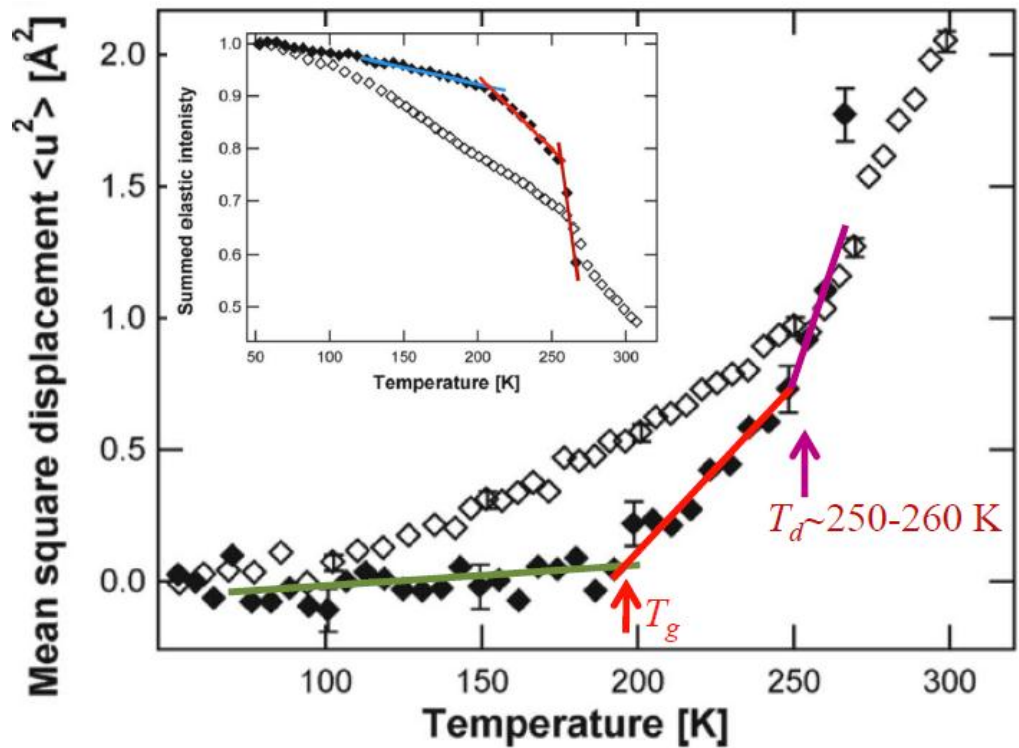

**Figure 6.** Mean square displacements (MSD) as a function of temperature obtained from elastic intensity on IN16 for D-PM-H$_2$O (filled diamonds) and H-PM-D$_2$O (open diamonds). Inset: Sum of the scattered normalized elastic intensity as a function of temperature for D-PM-H$_2$O (filled diamonds) and H-PM-D$_2$O (open diamonds). Reproduced from Ref.$^{29,51}$ by permission. The lines not in the original figures serve no other purpose except to suggest the presence of two breaks in *T*-dependence of the MSD for D-PM-H$_2$O, one at $T_g$~200 K and another one at $T_d$~250-260 K.

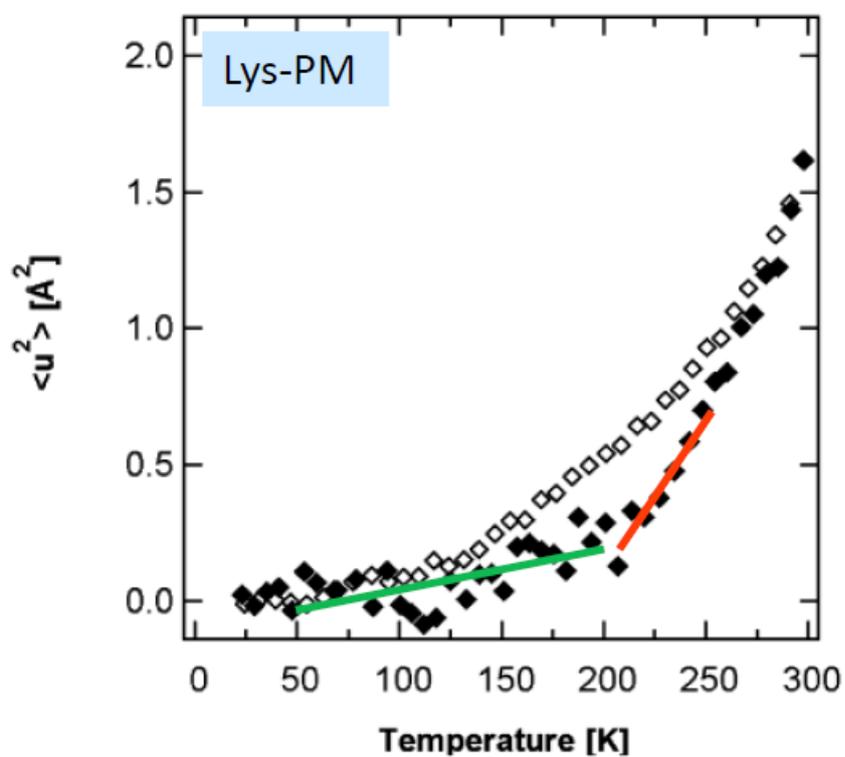

**Figure 7.** MSD from neutron scattering measurements on IN16 of natural abundance PM (open diamonds), and deuterated PM with hydrogenated lysine residues (filled diamonds). Reproduced from Ref.[26] by permission. The lines not in the original figures serve no other purpose except to suggest the presence of break in $T$-dependence for the deuterated PM with hydrogenated lysine residues at $T_g \sim 200$ K.